\documentclass[
journal=jacsat,
manuscript=communication,
amsmath,amssymb,
]{achemso}

\usepackage{chemformula} 
\usepackage[T1]{fontenc} 

\usepackage{natbib}

\usepackage{graphicx}
\graphicspath{{./}}

\usepackage{color}
\definecolor{darkblue}{rgb}{0,0,.8}
\usepackage{hyperref}
\hypersetup{pdftex=true, colorlinks=true, breaklinks=true, linkcolor=darkblue, citecolor=darkblue, menucolor=darkblue, pagecolor=darkblue, urlcolor=blue}

\usepackage[english]{babel}

\author{F. Dei\ss enbeck}
\affiliation[MPIE]
{Max-Planck-Institut f\"ur Eisenforschung GmbH, Max-Planck-Stra{\ss}e 1, 40237 D\"usseldorf, Germany}

\author{S. Wippermann}
\email{stefan.wippermann@physik.uni-marburg.de}
\affiliation[MPIE]
{Max-Planck-Institut f\"ur Eisenforschung GmbH, Max-Planck-Stra{\ss}e 1, 40237 D\"usseldorf, Germany}
\alsoaffiliation[UMR]
{Philipps-Universit\"at Marburg, Renthof 5, 35032 Marburg, Germany}

\title[AITP]
{Dielectric properties of nanoconfined water from \emph{ab initio} thermopotentiostat molecular dynamics}

\begin{document}

\begin{tocentry}
\center
\includegraphics[width=0.5\textwidth]{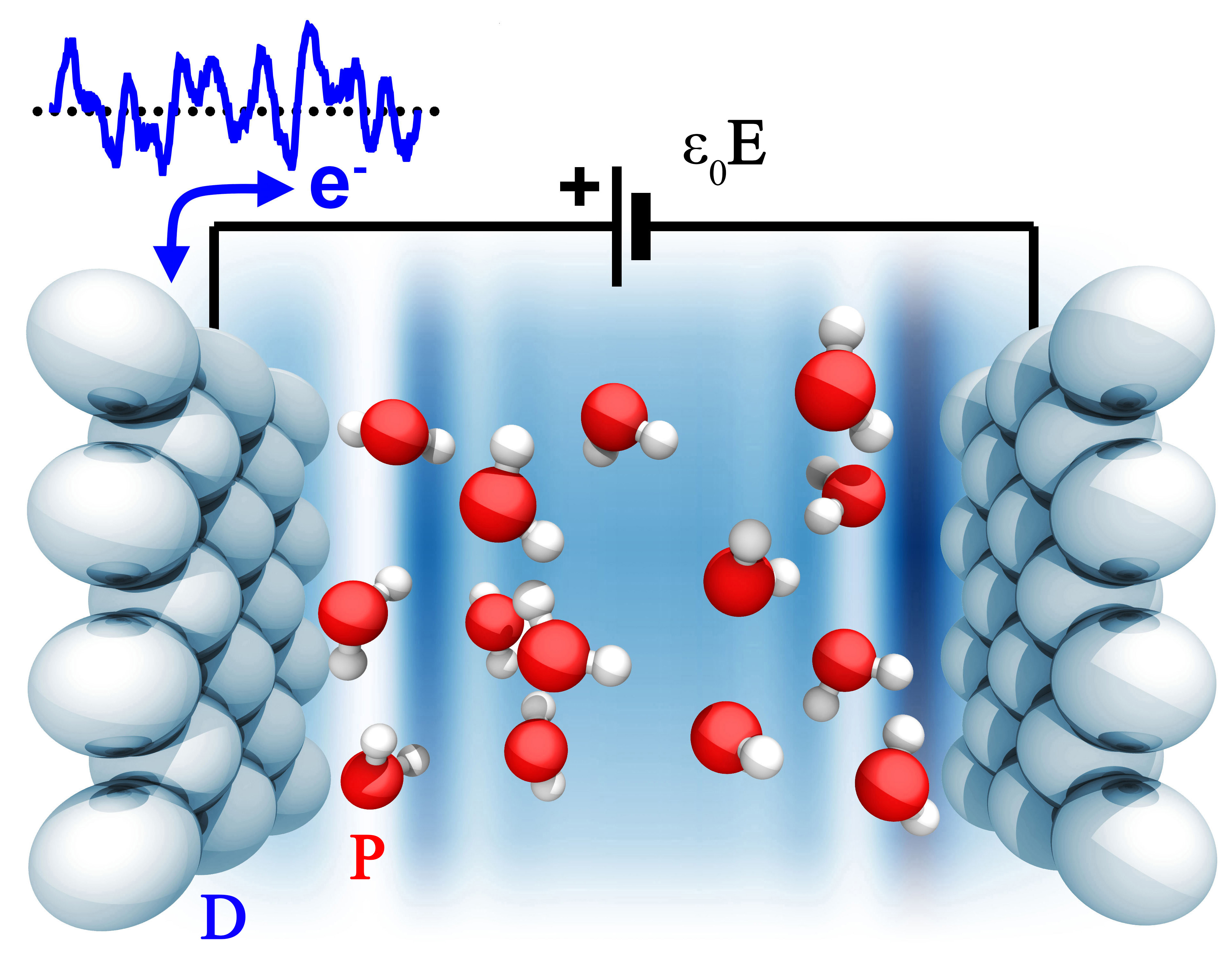}
\end{tocentry}

\begin{abstract}
We discuss how to include our recently proposed thermopotentiostat technique [Phys. Rev. Lett. \textbf{126}, 136803 (2021)] into any existing \emph{ab initio} molecular dynamics (AIMD) package. Using thermopotentiostat AIMD simulations in the canonical NVT$\Phi$ ensemble at constant electrode potential, we compute the polarization bound charge and dielectric response of interfacial water from first principles.
\end{abstract}


Electrochemical processes occurring at the interface between a solid electrode and an aqueous electrolyte are central to future sustainable energy conversion and storage solutions \cite{ciamician,jaramillo}. Applying a voltage across these interfaces enables control over the reaction pathways and kinetics. At electrified interfaces, however, water forms stratified layers \cite{marxrev} with properties that differ strongly from those observed in bulk solutions \cite{fumagalli}. Therefore, solvent reorganization in response to the electric field, ion de-/resolvation processes, the formation of the electric double layer and charge transfer reactions all proceed within these interfacial water regions with modified properties.

In order to enable further advances in these fields it is critical to develop accurate simulation techniques to explore and predict structural properties and chemical reactions at electrified surfaces in contact with liquid electrolytes from first principles. While experiments are routinely performed at constant electrode potential, realizing these conditions in \emph{ab initio} molecular dynamics (AIMD) simulations has remained very challenging. A suitable AIMD potentiostat technique requires two constituents: (i) a robust method to either apply an electric field or charge the electrode, and (ii) an algorithm to control either the field or charge in accordance with thermodynamic theory in order to drive the system to the desired electrode potential.

Multiple solutions have been suggested for issue (i). The modern theory of polarization (MTP) \cite{resta,stengel} explicitly includes the field inside the simulation cell, that is generated by moving the corresponding charge from one boundary of the unit cell to the opposite one. The charge itself is outside the unit cell. This approach has been used to perform first principles calculations with either a constant total electric field $\mathbf{E}$ \cite{stengel} or a constant electric displacement $\mathbf{D}$ \cite{stengel2} as electrostatic boundary conditions. Alternatively, Lozovoi and Alavi proposed to perform constant potential calculations by including an explicit compensating counter charge inside the unit cell to ensure charge neutrality \cite{lozovoi}. More recent approaches build on these ideas to either apply directly an electric field \cite{pasquarello,saitta,tavernelli,zhang,sayer,dufils,yang} or include explicit compensating counter charges \cite{esm,greens,frenzel,ashton,surendralal,honkala,selloni}.

We note that without exception, the techniques outlined above rely on electrostatic boundary conditions that enforce that either the total electric field $\mathbf{E}$ or the electric displacement $\mathbf{D}$ are kept exactly constant during the AIMD run. In the thermodynamic sense, all these methods describe purely microcanonical ensembles with constant total energy. In the thermodynamic limit, averaged properties are independent of the chosen ensemble. If, however, the observables of interest explicitly depend on fluctuations (e.g. reaction mechanisms and rates, etc.) or if the simulated system is small, the microcanonical ensemble may not be used. Instead, the electrode charge must be treated as a thermodynamic degree of freedom, allowing it to react to the dynamics of the solvent and to charge transfer processes. Such a treatment requires canonical sampling (issue ii).

The first such approach was pioneered by Bonnet \emph{et al.} \cite{otani} and later extended by Bouzid \emph{et al.} \cite{bouzid1,bouzid2}. Bonnet \emph{et al.} suggested to describe the electrode charge by second order dynamics coupled to a Nos\'e-Hoover thermostat. This approach, however, requires ''\emph{[...] the existence of an energy function $\mathcal{E}(r_i,n_e)$ that is differentiable with respect to the total electronic charge. This implies the ability to treat non-integer numbers of electrons and, in general, non-neutral systems.}`` \cite{otani}. Unfortunately, in the context of density-functional calculations the total energy as a function of the number of electrons is a notoriously difficult quantity to compute. Furthermore, the electronic charge is a single degree of freedom. Yet, controlling single degrees of freedom by the Nos\'e-Hoover method often leads to non-ergodic behaviour. In order to recover ergodicity, the introduction of Nos\'e-Hoover chains was proposed \cite{nhchain}, however, at the cost of additional numerical parameters and required extra tuning. In order to lift these requirements and enable a straightforward implementation of the potentiostating process into any simulation package, we were recently inspired by the MTP \cite{resta} and the Maxwell-Langevin equations of fluctuation electrodynamics \cite{rytov} to introduce a stochastic canonical thermopotentiostat algorithm \cite{deissenbeck}.

Here we discuss the implementation of our thermopotentiostat technique in the context of electronic structure calculations and \emph{ab initio} molecular dynamics. Specifically, we choose to build our implementation on the computational counter electrode (CCE) recently proposed by Surendralal \emph{et al.} \cite{surendralal}. In contrast to the finite field methods described in Refs. \cite{resta,stengel,stengel2,lozovoi,pasquarello,saitta,tavernelli,zhang,sayer,dufils,yang,esm,greens,frenzel,ashton,honkala,selloni}, which are available in only some of the most commonly used density-functional theory (DFT) codes, the CCE technique has the added advantage that its application does not require any changes inside the electronic structure code. However, we emphasize that the thermopotentiostat algorithm is equally straightforward to implement using any of the methods outlined in Refs. \cite{resta,stengel,stengel2,lozovoi,pasquarello,saitta,tavernelli,zhang,sayer,dufils,yang,esm,greens,frenzel,ashton,honkala,selloni}.

\begin{figure}[t]
  \centering
    \includegraphics[width=0.48\textwidth]{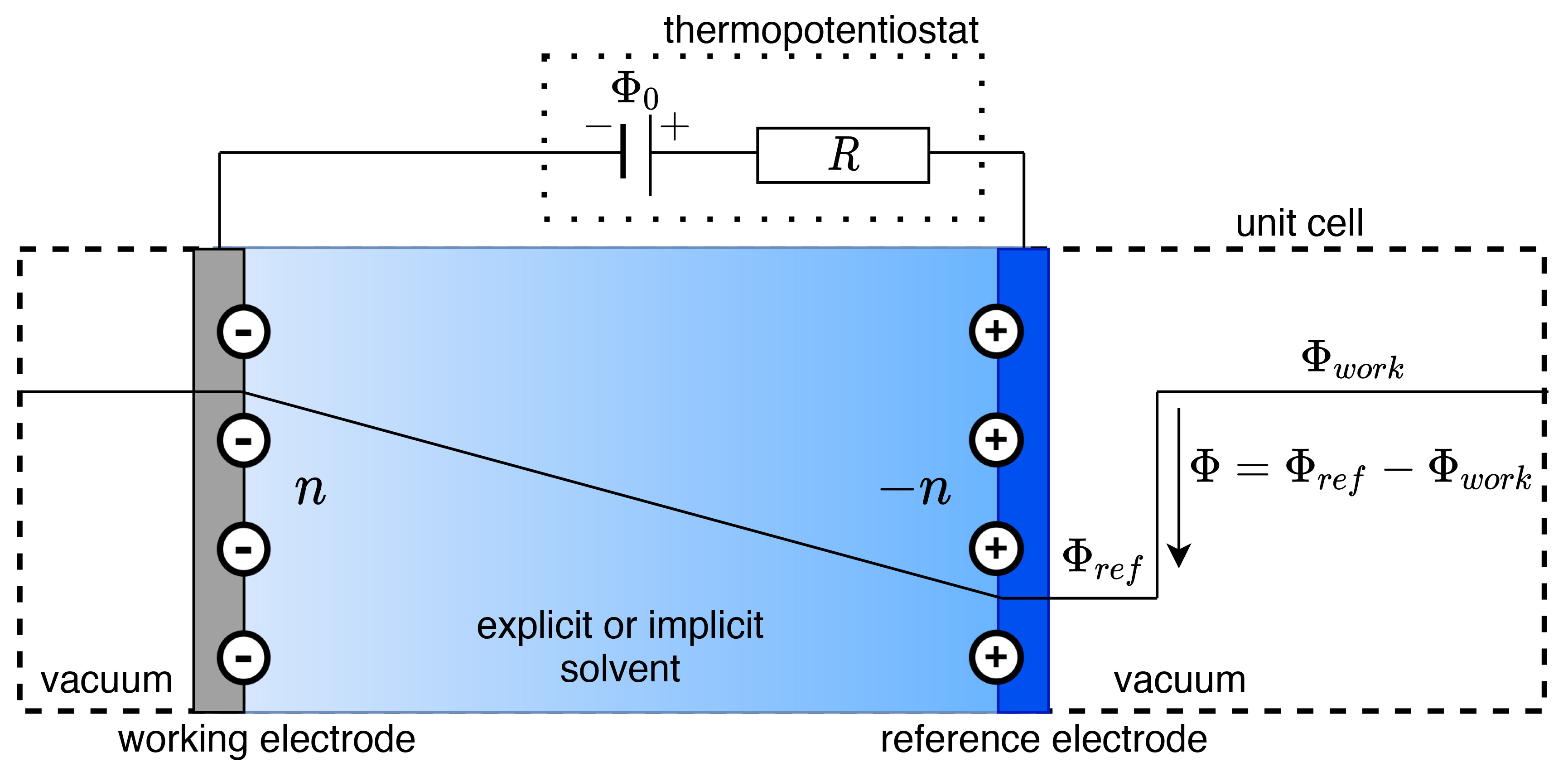}
\caption{\label{definitionphi} Schematic representation of the computational setup. The periodic simulation cell is indicated by the dashed line. The power supply and resistor located outside the unit cell represent the thermopotentiostat introduced in Ref. \cite{deissenbeck}. As an input quantity, the thermopotentiostat requires the instantaneous potential $\Phi$. It is determined from the difference of the workfunctions on the vacuum side and solvent covered surface of the working electrode, and it is equal to the total dipole moment of the charges contained in the simulation cell along the surface normal. The potential as drawn corresponds to the electron potential, consistent with the definition commonly used in electronic structure codes. By convention, we choose that an increasingly positive potential corresponds to an increasingly positive charge on the working electrode.}
\end{figure}

Fig. \ref{definitionphi} illustrates the computational setup chosen in the present study. The simulation cell contains an electrolyte or dielectric medium (explicit or implicit) that is enclosed between a working and a reference electrode, carrying equal and opposite charges $n$ and $-n$, respectively. Hence, the simulation cell is charge-neutral in total. The working electrode is connected to an external reservoir of charge at constant electron chemical potential, so that the \emph{external} voltage difference $\Phi_0$ between the working and reference electrodes is exactly constant. The potential $\Phi_0$ is the independent thermodynamic variable that can be controlled in experiments. The \emph{system} potential $\Phi$ inside the simulation cell, in contrast, is defined as the difference of the workfunctions on the vacuum side and the solvated surface of the working electrode, cf. Fig. \ref{definitionphi}. It is neither a constant nor necessarily equal to the bath potential, due to the microscopic size of the region targeted by our simulations and the exchange of charge with the external environment. Instead, the system potential $\Phi$ as well as the Fermi level and the charge $n$ depend on the evolution of the system.

Controlling the charge $n$ at each discrete simulation time step allows us, in principle, to drive the system potential $\Phi$ towards the desired target value for the external bath potential $\Phi_0$. Treating thereby the system potential as a thermodynamic degree of freedom implies, however, that the simulation is able to perform external work and, hence, dissipate energy. In order to uphold energy conservation, the energy loss due to controlling the system potential must be balanced exactly, on average, by a corresponding energy gain from thermally induced fluctuations. The physical system realises this condition by coupling to a fluctuating electric field created by temperature dependent charge fluctuations due to the thermal motion of the electrons and ions \cite{callen}. To mimic this behaviour, a potentiostat must apply an electric field with an explicit finite temperature and hence become a ``thermopotentiostat''. We introduced such an algorithm in Ref. \cite{deissenbeck} and derived a direct expression for the electrode charge $n$ at each discrete time step:
\begin{eqnarray}
n(t + \Delta t) = n(t) &-& \underbrace{C_0 \left(\Phi(t) - \Phi_0 \right) \left(1-e^{-\frac{\Delta t}{\tau_{\Phi}}} \right)}_{dissipation} \nonumber\\ &+& \underbrace{N \sqrt{k_B TC_0 \left(1 - e^{-\frac{2 \Delta t}{\tau_{\Phi}}}\right)}}_{fluctuation}, \label{ft}
\end{eqnarray}
where $n$ is the electrode charge and $N$ is a random number drawn from a Gaussian distribution with zero mean and variance one. $C_0$ is the geometric capacitance of the bare electrodes in absence of the dielectric or solvent and $\tau_{\Phi} := R C_0$ is the potentiostat relaxation time constant. The instantaneous system potential $\Phi(t)$ of the working electrode with respect to the reference electrode is obtained from the total dipole moment of the charges contained in the simulation cell parallel to the surface normal \cite{deissenbeck}. In practice, $\Phi(t)$ is computed using the dipole correction scheme \cite{neugebauer1992adsorbate} that is available in most density-functional codes. We use the convention that an increase of the potential in the positive direction implies an increasingly anodic polarization on the working electrode, indicated by the vertical arrow in Fig. \ref{definitionphi}.

We note that Eq. \ref{ft} determines only the total amount of charge $n$ present on the working electrode at each discrete time step. In the context of Born-Oppenheimer (BO) DFT, the actual distribution of $n$ on the electrode surface is determined by the electronic minimization. However, since in BO approximation there is no explicit electron dynamics and hence no scattering, the electrode charge is redistributed instantaneously each time step effectively describing an electrode with infinite surface conductivity. Therefore, the physical meaning of the resistance $R$ shown in Fig. \ref{definitionphi} -- and by extension the relaxation time $\tau_{\Phi}$ in Eq. \ref{ft} -- is to introduce an effective mean surface conductance that governs the flow of charge into and out of the finite segment of the electrode described within the simulation cell.

In practice, $\tau_{\Phi}$ is set to a sufficiently small value to enable an efficient sampling of the phase-space but not yet small enough to disturb the system dynamics. Note that the mean and the variance of the charge as given by Eq. \ref{ft} are unaffected by the choice of $\tau_{\Phi}$. Small values of $\tau_{\Phi}$, however, correspond to a large damping factor and may thus adversely affect the dynamics of the system if set too aggressively. In general, a time-constant longer than the slowest vibrational frequency present in the system is a reasonable choice. We therefore adopt $\tau_{\Phi} = 100$ fs as a default value.

Moreover, note that the present first principles computational setup differs from the semi-empirical one described in Ref \cite{deissenbeck} in an important aspect: in BO-DFT calculations, the simulated system is instantaneously polarizable. Since the capacitance enters Eq. \ref{ft} as an adjustable parameter, the potentiostating process seemingly requires prior knowledge about the dielectric properties of the system. This, however, is not the case. In fact, Eq. \ref{ft} already takes the instantaneous polarizability correctly into account. This property of Eq. \ref{ft} can be understood intuitively, considering that the instantaneous electric current $\frac{d n}{dt}$ is independent of the capacitance. It is only the discrete change $n(t + \Delta t) - n(t)$ that depends on the capacitance. For convenience, the construction of Eq. \ref{ft} treats any deviations in the actual capacitance $C = \epsilon_r C_0$ with respect to the parameter $C_0$ within the time domain: even if $\epsilon_r$ contains a contribution due to instantaneous polarizability, Eq. \ref{ft} is guaranteed to sample the correct statistical distribution for the charge $n$ with $\sigma_n^{2} = k_B T \epsilon_r C_0$, albeit with an increased relaxation time of $\epsilon_r\tau_{\Phi}$. An analytical proof of this somewhat counter-intuitive property of Eq. \ref{ft} is included in the SI \cite{supp}.

Conceptually, the parameter $C_0$ plays a role analogous to the mixing parameter $\beta$ of the charge mixing schemes \cite{mixBroyden,mixPulay,mixJohnson,mixKerker} commonly used in DFT, where $\beta$ determines how much of the old density is mixed to the new density from one electronic iteration to the next. In case one of the dimensions of the unit cell is significantly larger than the other two, small changes of the electron density with respect to the more extended direction are associated with large changes in the total energy. In essence, such a unit cell corresponds to a reduced capacitance $C_0$, which scales with $1/l$, where $l$ is the length of the unit cell. Thus, in order to prevent charge sloshing and convergence issues it is often necessary to reduce the mixing parameter $\beta$ in these situations. The parameter $C_0$, like $\beta$, is of purely numerical nature and has no impact on any physical observable. However, it must be chosen appropriately to ensure numerical stability. The exact value of $C_0$ is uncritical and its choice is straightforward: setting $C_0$ to approximately $\epsilon_0 A/d$, where $A$ is the area of the unit cell parallel to the electrode surface and $d$ is the distance between the working and the reference electrodes resulted in stable convergence behaviour in all cases investigated.

We now turn to discuss the implementation of our thermopotentiostat into existing AIMD packages. The implementation must be built on top of a method to realise either a finite charge on the working electrode or apply a finite electric field. The thermopotentiostat is completely general and can be used to control either the field or charge in conjunction with any of the methods described in Refs. \cite{resta,stengel,stengel2,lozovoi,pasquarello,saitta,tavernelli,zhang,sayer,dufils,yang,esm,greens,frenzel,ashton,surendralal,honkala,selloni}. As a basis for our implementation, here we chose the computational counter electrode (CCE) recently proposed by Surendralal \emph{et al.} \cite{surendralal}, because the CCE can be directly used with any DFT code. Building on the CCE, only the thermopotentiostat needs to be implemented inside the electronic structure code as a control scheme in analogy to a thermostat, but not the finite field method itself.

In their scheme, Surendralal \emph{et al.} \cite{surendralal} used a large band gap insulator, so that the Fermi level of the total system can be controlled within the electronic gap of the CCE by doping. To transfer charge between the working electrode and the CCE, Surendralal \emph{et al.} suggested to dope the CCE using pseudoatoms with fractional core charges $Z_{CCE}$. This approach places an adjustable charge on the working electrode and at the same time provides an equal and opposite compensating counter charge on the CCE.

In the following, we couple the thermopotentiostat to the CCE: at each ionic step, the thermopotentiostat is used to determine the change of the charge $n$ that is located on the segment of the working electrode described within the simulation cell, exchanged with an external bath at constant electron chemical potential. The new electrode charge is then realised by adjusting the core charge of the atoms constituting the CCE according to:

\begin{equation}
Z_{CCE}(t + \Delta t) = Z_{CCE}(t) + \frac{n(t + \Delta t) - n(t)}{N_{CCE}}, \label{zne}
\end{equation}
where $N_{CCE}$ is the number of atoms constituting the counter electrode and $n$ is computed according to Eq. \ref{ft}. 

\begin{figure}[t]
  \centering
    \includegraphics[width=0.4\textwidth]{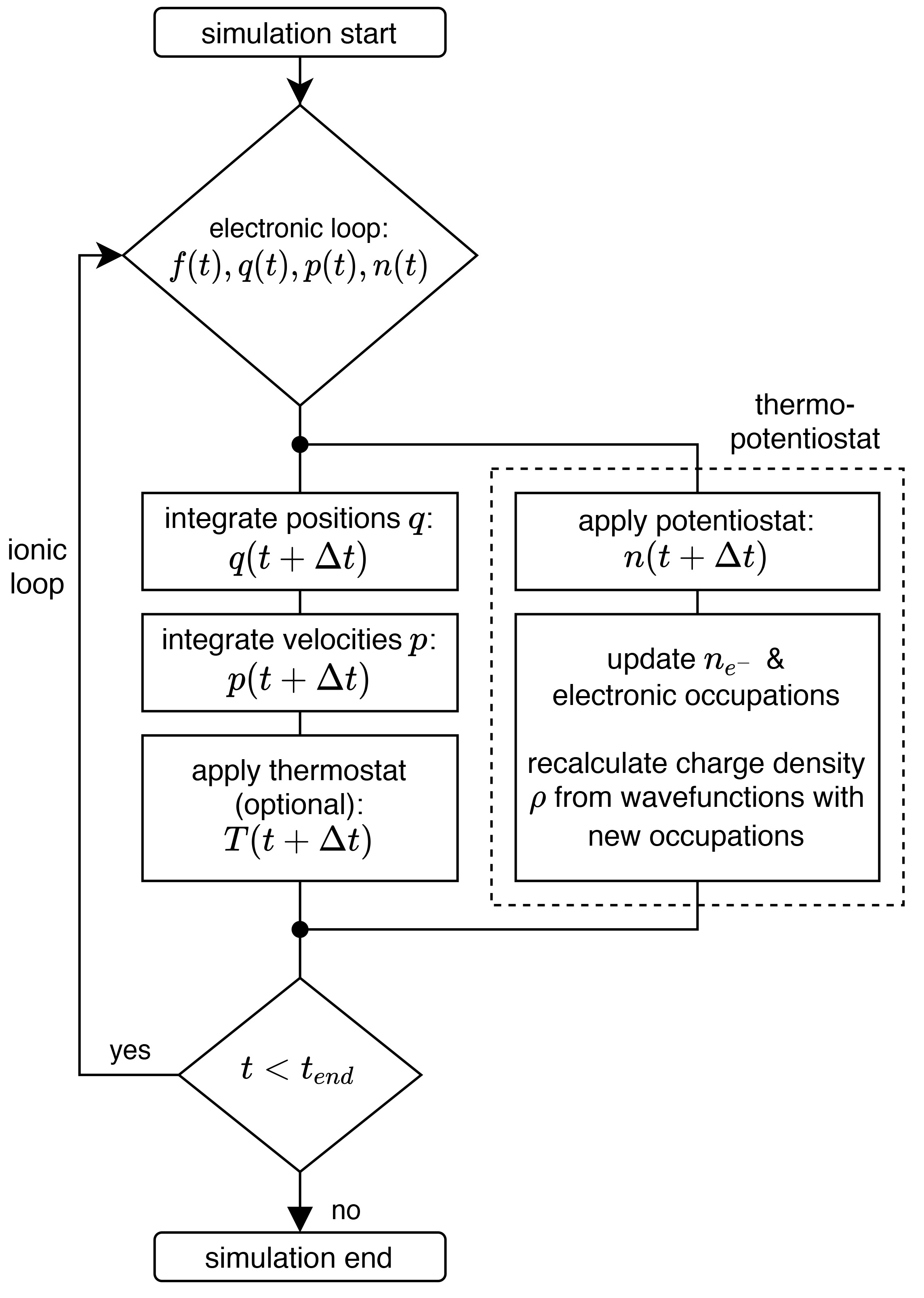}
\caption{\label{integrationscheme} Flowchart of the thermopotentiostat implementation in conjunction with leapfrog integration of the equations of motion, see text. A corresponding flowchart for velocity Verlet integration is included in the SI \cite{supp}.}
\end{figure}

Most BO-DFT codes rely either on the (velocity) Verlet or leapfrog algorithms to integrate the equations of motion. Fig. \ref{integrationscheme} outlines a code structure commonly used throughout many existing DFT packages, complemented by our thermopotentiostat. After the calculation of the electronic structure and the forces at time $t$, both the positions and the electrode charge (or $Z_{CCE}$, respectively) can be updated directly and in any order to time $t + \Delta t$. The code structure differs slightly if velocity Verlet integration is used, cf. Fig. S1 \cite{supp} for a representative example.

We note that although our computational setup guarantees that the total system is always charge neutral, the number of electrons $n_{e^-}$ contained inside the simulation cell is free to change from one ionic step to the next, as the thermopotentiostat adjusts the compensating counter charge located on the CCE. After computing the updated electrode charge $n(t + \Delta t)$ and doping the CCE according to Eq. \ref{zne}, it is therefore necessary to update also the number of electrons $n_{e^-}$ perceived by the electronic structure code. If the number of electrons changes, by default, most DFT codes will shift the electron density by a constant offset so that its volume integral becomes equal to the new electron number. A straightforward shift of the electron density, however, may cause charge sloshing during the scf cycle and led to convergence issues in previous approaches \cite{lozovoi}. The presence of explicit fluctuations in our approach exacerbates this problem further.

To ensure that the electronic loop converges reliably, we note that in the physical system charge is added to or removed from the electrode at the Fermi level only. In order to recover the physically correct behaviour in our setup, we recalculate the electronic occupations for the electronic structure at time $t$, but with the already updated number of electrons $n_{e^-}(t + \Delta t)$. Subsequently, the electron density $\rho$ is recalculated from the present wavefunctions at time $t$ using the new occupations, and introduced into the Hamiltonian at time $t+\Delta t$. We tested this approach at the example of different semiconducting and metallic working electrodes, and we obtained a completely robust electronic convergence behaviour in all cases investigated.

In order to highlight the opportunities provided by our \emph{ab initio} thermopotentiostat technique, we now turn to a topic that currently attracts considerable attention: interfacial water and water under confinement feature structural and dynamic properties that differ significantly from those of bulk water \cite{marxrev}. Most notably, thin water films confined to a few nm thickness exhibit a strongly reduced dielectric response in the direction perpendicular to the confining surfaces \cite{fumagalli}. Since the high polarizability of liquid water is regarded as the origin of its unique solvation behaviour and interfacial water is omnipresent, it is necessary to understand the mechanism and to be able to accurately simulate the dielectric response of liquid water at realistic interfaces.

Fumagalli \emph{et al.}'s work \cite{fumagalli} therefore stimulated a considerable number of theoretical studies, cf. e.g. Refs. \cite{marxpccp,loche,schlaich2,motevas,matyushov,laage}. Most studies used Kirkwood-Fr\"ohlich theory \cite{sharma,neumann,kirkwood} or the theory of polarization fluctuations \cite{feller} in order to compute the dielectric tensor from the variance of the total dipole moment fluctuations per volume. Converging these variances enforces, however, a statistical sampling of the water dynamics spanning a time scale of several hundred ns. For this reason, atomistic simulations of interfacial water's dielectric response have been largely restricted to force-field approaches and non-polarizable water models.

Moreover, these simulations require the dielectric volume as an input parameter, e.g. in the form of dielectric box models \cite{schlaich2}. The dielectric volume is well defined for homogeneous bulk media. In the presence of an interface between the electrode and the dielectric, however, the exact location of the boundary and hence the definition of the dielectric volume become ambiguous. This ambiguity reflects an assumption implicit to the construction of both the Kirkwood-Fr\"ohlich theory and the theory of polarization fluctuations: these approaches describe the dielectric response of the system enclosed within the periodic simulation cell to a displacement field created by sheet charges that are placed exactly on the boundaries of the simulation cell. In atomistic simulations of electrified interfaces, however, the distribution of the electrode charge may differ significantly from idealized sheet charges, which Kirkwood-Fr\"ohlich and the theory of polarization fluctuations are fundamentally unable to account for.

Both problems are solved by our thermopotentiostat technique in conjunction with explicit applied fields. Introducing the densities of the electrode charge $n$ and the bound charge $n_p$ as the explicit quantities to describe the response of a dielectric medium exposed to an electric field enables direct and parameter-free access to dielectric properties. Moreover, the densities can be computed at least two orders of magnitude faster than dipole variances, due to the use of thermodynamic averages and the efficient stochastic canonical sampling of our approach. Themopotentiostat MD thereby opens the door towards simulations of interfacial dielectric properties from first principles.

\begin{figure}[t]
  \centering
    \includegraphics[width=0.48\textwidth]{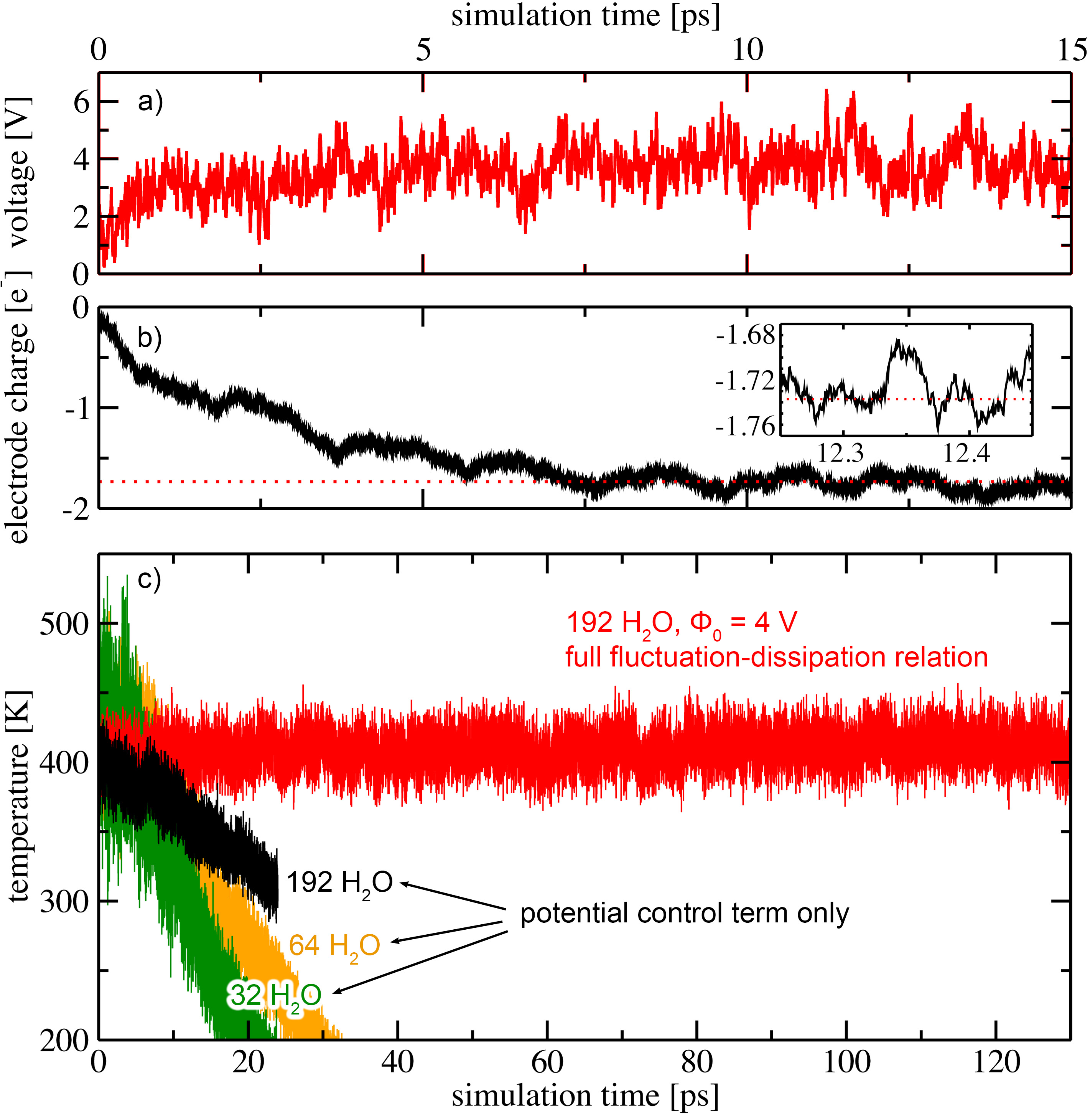}
\caption{\label{potentiostat} 
\textbf{a)} Time evolution of the electrode charge $n$ and \textbf{b)} the system potential $\Phi$ for an NVE ensemble consisting of 192 H$_2$O molecules, potentiostated to $\Phi_0 = 4$ V. The red dotted line marks the average electrode charge after equilibration. The inset shows the stochastic charge fluctuations in an enlarged region around the average electrode charge. \textbf{c)} Time evolution of the temperature for multiple NVE ensembles consisting of 32, 64 and 192 H$_2$O molecules, respectively, potentiostated to $\Phi_0 = 4$ V. The electrode charge is adjusted, using only the dissipation (potential control) term in Eq. \ref{ft} (green, orange and black lines), or using the full Eq. \ref{ft} (red line).
}
\end{figure}

We implemented our thermopotentiostat approach into the Vienna Ab Initio Simulation Package (VASP) and performed AIMD simulations for liquid water confined between two computational Ne electrodes, using the generalized gradient approximation PBE \cite{vasp1, vasp2}. As test cases we considered ensembles consisting of $32$, $64$ and $192$ water molecules, corresponding to electrode separations of $d=10.7$, $17.4$ and $31.4$ \r{A}, respectively. Consistent with the work of Fumagalli \emph{et al.} \cite{fumagalli}, we applied potentials of $\Phi_0 = 0$ V and $4$ V, respectively. Further numerical details are provided in the SI \cite{supp}.

In Fig. \ref{potentiostat}a and b we show the time evolution of the potential difference and the charge transferred between the two electrodes, respectively, directly after switching on the thermopotentiostat. The system potential is driven efficiently towards the externally applied voltage and becomes stationary after a simulation time of $\sim 4$ ps has elapsed. Yet, a net current continues to flow until $\sim t = 8$ ps, where the charge assumes an equilibrium value of $\bar{n} = -1.73\ e^-$, cf. Fig. \ref{potentiostat}b. This electric displacement current is caused by the increasing polarization density of the water due to the reorientation of the water molecules within the applied electric field. This reorientation occurs on a much slower time scale than the changes in the applied electric field, which are governed by the chosen relaxation time constant of $\tau_{\Phi} = 100$ fs. Moreover, after the system has reached equilibrium the electrode charge continues to fluctuate (inset Fig. \ref{potentiostat}a). We note that two different processes contribute to these fluctuations: on the one hand, any water dynamics that are associated with a change in the total dipole moment, as well as contributions due to electronic screening, are actively countered by the thermopotentiostat in order to drive the potential towards its target value. This action of the thermopotentiostat therefore takes the form of deterministic fluctuations in the electrode charge. On the other hand, the energy dissipated due to potential control is returned, on average, by the stochastic fluctuations introduced in Eq. \ref{ft}.

An important aspect of simulations under potential control is to ensure that the interplay between the deterministic potential control mechanism and the stochastic fluctuations balances out to a zero net energy change, in order to sample the canonical ensemble at constant temperature and applied potential. As discussed above and in the supplemental information \cite{supp}, our central working Eq. \ref{ft} explicitly takes instantaneously polarizable systems into account. For verification, we performed all simulations presented here in absence of an explicit thermostat. Neglecting the fluctuation term in Eq. \ref{ft}, the potential control mechanism alone always dissipates thermal energy from those vibrational degrees of freedom, that couple to a change in the total dipole moment of the ensemble. For typical DFT system sizes the pure potential control mechanism is able to drive the ensemble severely out of equilibrium in a matter of ps, cf. Fig. \ref{potentiostat}c (black, orange and green curves, respectively). If, in contrast, the full fluctuation-dissipation relation Eq. \ref{ft} is used (cf. Fig. \ref{potentiostat}c, red curve), the thermopotentiostat eliminates any artificial net energy drain and the temperature remained constant, on average, over the whole course of the simulation. 

\begin{figure}[t]
  \centering
    \includegraphics[width=0.48\textwidth]{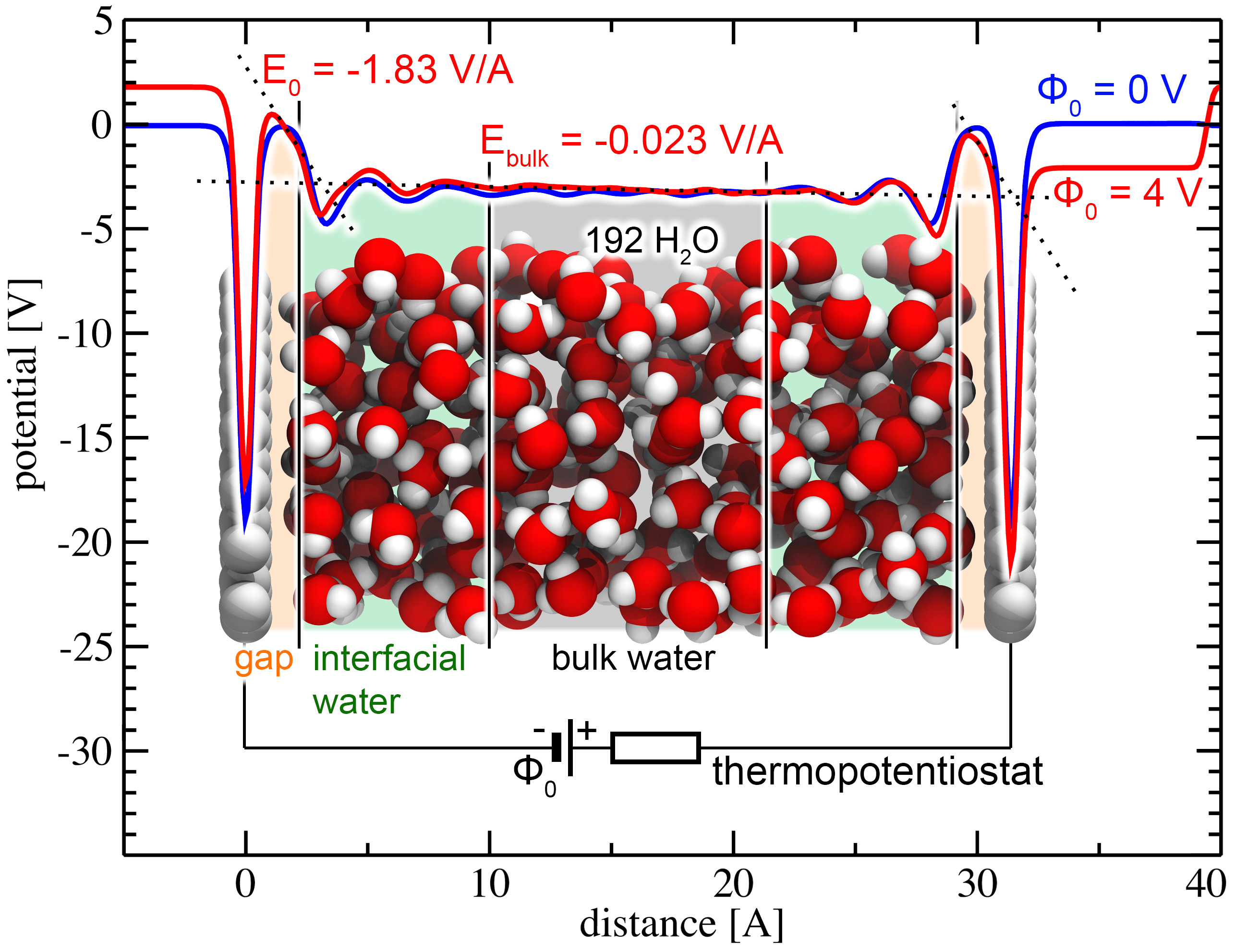}
\caption{\label{dftsetup}
Schematic representation of the \emph{ab initio} simulation cell. Grey balls represent electrode surface atoms (Ne), whereas red and white balls denote O and H, respectively. The unit cell has a lateral dimension of 14.5$\times$14.5 \AA$^2$, with a distance of $d = 31.4$ \AA\ between the electrodes, and contains 192 H$_2$O molecules at the experimental density of water. Blue and red lines indicate planar-averaged electrostatic potential profiles parallel to the surface normal for applied voltages of $\Phi_0 = 0$ V and $\Phi_0 = 4$ V, respectively, time-averaged over a trajectory length of $125$ ps, see text.
}
\end{figure}

We now explore the dielectric properties of nanoconfined water, using our thermopotentiostat technique to include the atomistic details of the interface at the AIMD level of theory. In Fig. \ref{dftsetup}, we compare electrostatic potential profiles that were obtained for two different external applied voltages $\Phi_0 = 0$ V (blue curve) and $\Phi_0 = 4$ V (red curve), respectively, each time-averaged over a trajectory length of $125$ ps. We partitioned the space between the electrodes into three different regions: (i) a hydrophobic gap with a thickness of 2 \AA, formed between the electrode surface and interfacial water, (ii) an interfacial water region with a thickness of 8 \AA, and (iii) a bulk-like water region, indicated in Fig. \ref{dftsetup} by areas shaded in orange, green and grey, respectively.

At the positions of the electrodes, the nuclear core charges of the atoms constituting the electrodes induce deep wells in the potential. For an applied voltage of $\Phi_0 = 4 V$, the potential then decays linearly and almost unscreened within the hydrophobic gap regions, resulting in a homogeneous electric field of $E_{0,\perp} = -1.83$ V/\AA. The electric field is also homogeneous and constant inside the bulk-like water region with a value of $E_{\mathrm{bulk},\perp} = -0.023$ V/\AA. The field strengths are indicated by dotted lines in Fig. \ref{dftsetup}. We then estimated the static dielectric constant inside the bulk-like water region as the ratio between the unscreened electric field within the hydrophobic gap and the field in the central bulk-like region. We obtained a value of $\epsilon_{\mathrm{bulk},\perp} = E_{0,\perp} / E_{\mathrm{bulk},\perp} \approx 79$, consistent with the one for homogeneous bulk water.

Although the dielectric response within the central bulk-like water region is fully consistent with the continuum picture shown in Fig. \ref{definitionphi}, the region of interfacial water exhibits a distinctly different behaviour. At interfaces, water forms stratified structures \cite{li2019situ, raghavan1991structure}. This stratification gives rise to potential oscillations within the interfacial water region, cf. green shaded area in Fig. \ref{dftsetup}. In analogy to Friedel oscillations \cite{friedel}, which originate when screening an electric field with charge carriers of finite size, the wavelength of the potential oscillations reflects the size of the water molecules \cite{waterfriedel}. In consequence, a considerable part of the potential drop applied between the two electrodes occurs within the hydrophobic gap, where the field is essentially unscreened, and inside the interfacial stratified water region. Since the gap and interfacial water regions are unable to effectively screen the applied electric field, the total static dielectric constant $\epsilon_{\perp}$, as measured by capacitive techniques \cite{fumagalli}, is significantly lower for nanoconfined water than that of homogeneous bulk water \cite{deissenbeck}.

The dielectric response of interfacial water is often characterized using spatially resolved dielectric constants and dielectric profiles \cite{loche,schlaich,schlaich2,motevas,marxpccp,deissenbeck}. This approach has recently been criticized on the grounds that physically meaningful dielectric constants can only be obtained at the mesoscale, averaging over multiple molecular layers \cite{laage}. Moreover, the ability of the interfacial water layer to polarize and store electrostatic free energy (dielectric response) should not be confused with the reduction of Coulomb interactions between charges, e.g. ions, embedded inside this region (screening). These two properties diverge at the nanoscale \cite{matyushov} and cannot be described by a spatially dependent local dielectric constant, due to the granularity of the solvent.

The response of a dielectric medium to electric fields or embedded charges, however, is well-defined even at the nanoscale in terms of the induced polarization density $\mathbf{P}$ and hence -- by extension -- the corresponding local bound charge density $\rho_{\mathrm{bound}}$. We therefore propose to use $\rho_{\mathrm{bound}}$ as the central local quantity in the context of first principles atomistic simulations.

In order to explicitly compute $\rho_{\mathrm{bound}}$, we begin by calculating the total charge density $\rho_{\mathrm{tot}} = - \epsilon_0 \cdot \Delta \phi$ from the DFT effective single particle potential $\phi(\mathbf{r})$, according to Poisson's equation \footnote{We note that most \emph{ab initio} codes define the electrostatic potential as the electron potential, which has the opposite sign compared to classical electrostatics. In Fig. \ref{dftsetup}a, we adopt the same convention, so that $\phi$ is equal to the inverse potential shown in Fig. \ref{dftsetup}a.}. Note that $\rho_{\mathrm{tot}}$ includes both the electronic charge as well as the nuclear core charge. Subsequently, $\rho_{\mathrm{tot}}$ is partitioned into a free charge contribution $\rho_{\mathrm{free}}$, whose volume integral is equal to the electrode charge $n$, and the bound charge contribution $\rho_{\mathrm{bound}}$. We computed the free charge density $\rho_{\mathrm{free}}$ using the identical 2-electrode setup but without the water as a reference. Of course other approaches to partition the charge density are conceivable as well, e.g. based on Wannier function techniques \cite{wannier,wannier_rev}. The bound charge density is then obtained as $\rho_{\mathrm{bound}} = \rho_{\mathrm{tot}} - \rho_{\mathrm{free}}$.

\begin{figure}[t]
  \centering
    \includegraphics[width=0.48\textwidth]{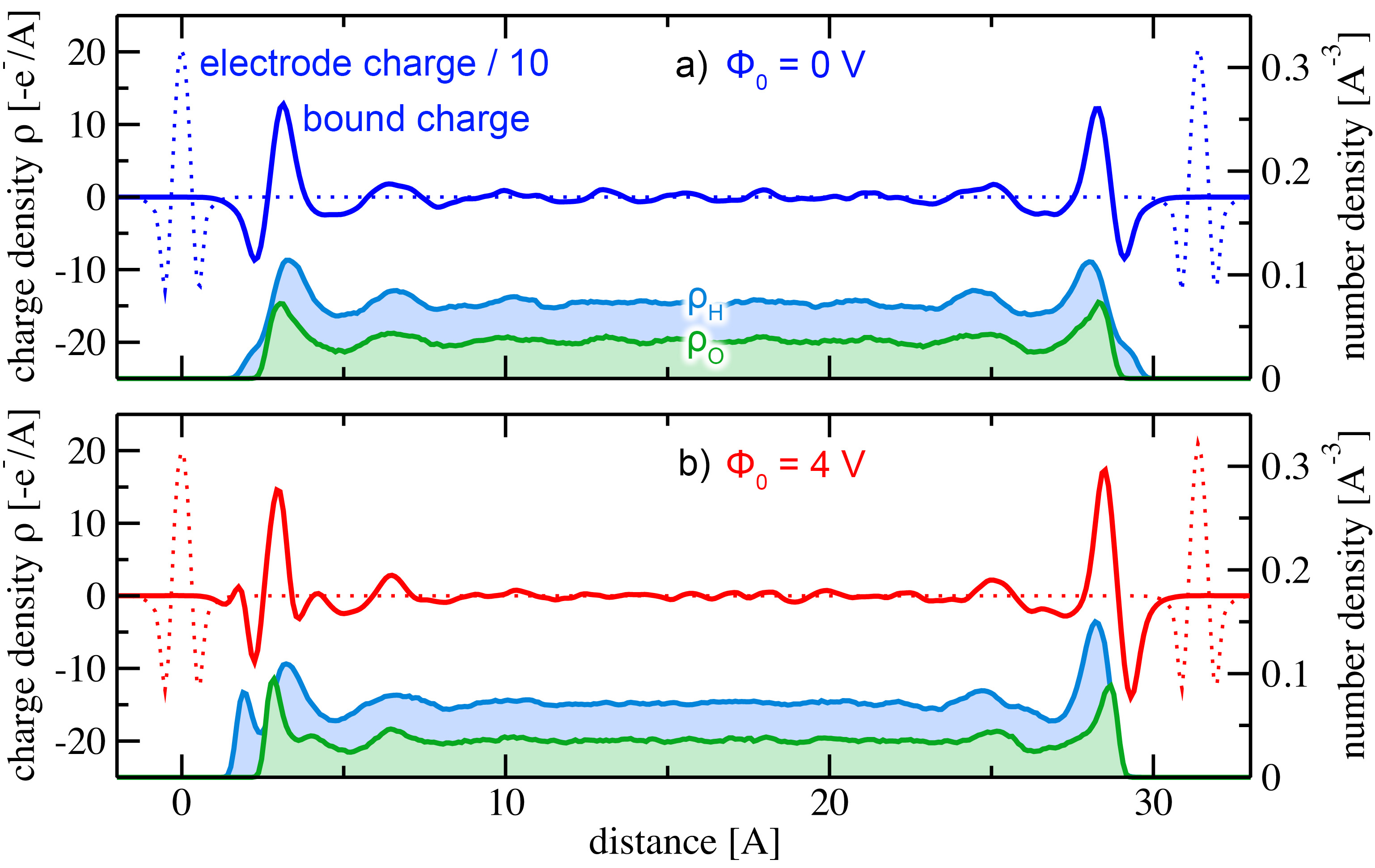}
\caption{\label{rhophi}
\textbf{a)} Bound charge density profiles for \mbox{$\Phi_0 = 0$ V} and \textbf{b)} \mbox{$\Phi_0 = 4$} V, obtained by subtracting the total charge density of the electrodes as a reference, see text. For improved visibility, the electrode charge densities shown as dotted lines have been scaled by a factor of 1/10. Filled light green and light blue areas indicate O/H number density profiles, respectively.
}
\end{figure}

Figs. \ref{rhophi}a and \ref{rhophi}b show line profiles of the resulting bound charge densities parallel to the surface normal for applied potentials of $\Phi_0 = 0$ V and $\Phi_0 = 4$ V, respectively. For both voltages, the bound charge density is zero inside the bulk-like water region to within our numerical accuracy. This is the expected outcome. Inside the regions of interfacial water, however, the bound charge density displays characteristic oscillations, even for zero applied voltage (cf. Fig. \ref{rhophi}a). These oscillations correspond to the specific structure assumed by water at the interface, in particular the stratification discussed above. In Fig. \ref{rhophi}a, water stratification is visible as the modulation in the water number density. The bound charge density hence gives rise to a characteristic dipole moment within the interfacial water layers. The exact details depend on the specific electrode and may include charge transfer due to e.g. chemisorbed water.

This purely \emph{chemical contribution} to the bound charge density is to be clearly distinguished from \emph{field-induced} contributions due to the presence of surface charges. For an applied voltage of $\Phi_0 = 4\ \mathrm{V}$, Fig. \ref{rhophi}b shows distinct modifications to the bound charge density. These modifications coincide with field-induced changes to the internal structure of the solid-water interface, as illustrated by the water number density shown in the bottom of Fig. \ref{rhophi}b. Since the left electrode is negatively charged, one of the hydrogen atoms of each interfacial water molecule is now pointing towards the electrode while the other one remains available to contribute to the hydrogen bond network. This reorientation is visible in the form of a double peak structure in the H number density close to the left hand interface. It is also reflected in the distribution of the angle between the OH-bond and the surface normal. A recent study \cite{li2019situ} reported similar findings for water-gold interfaces. We include the calculated angle distributions in the SI \cite{supp}.

In principle, both the \emph{chemical} and \emph{dielectric response} of a solvent to a given solute can be accurately described in terms of the bound charge densities discussed above. Bound charge densities computed from first principles, hence, represent important benchmarking quantities that allow us to evaluate the performance of implicit solvent models, such as e.g. modified Poisson-Boltzmann (MPB), models based on the integral theory of liquids (e.g. RISM) and molecular DFT \cite{rev_reuter}.

Despite the above mentioned shortcomings \cite{laage,matyushov} of describing the nanoscale dielectric response using local dielectric constants, this approach is often desirable for practical reasons in the construction of implicit solvent models. There has hence been continued interest to compute local dielectric profiles \cite{loche,schlaich,schlaich2,motevas,marxpccp,deissenbeck,marxrev} for water-solid interfaces.

\begin{figure}[t]
  \centering
    \includegraphics[width=0.48\textwidth]{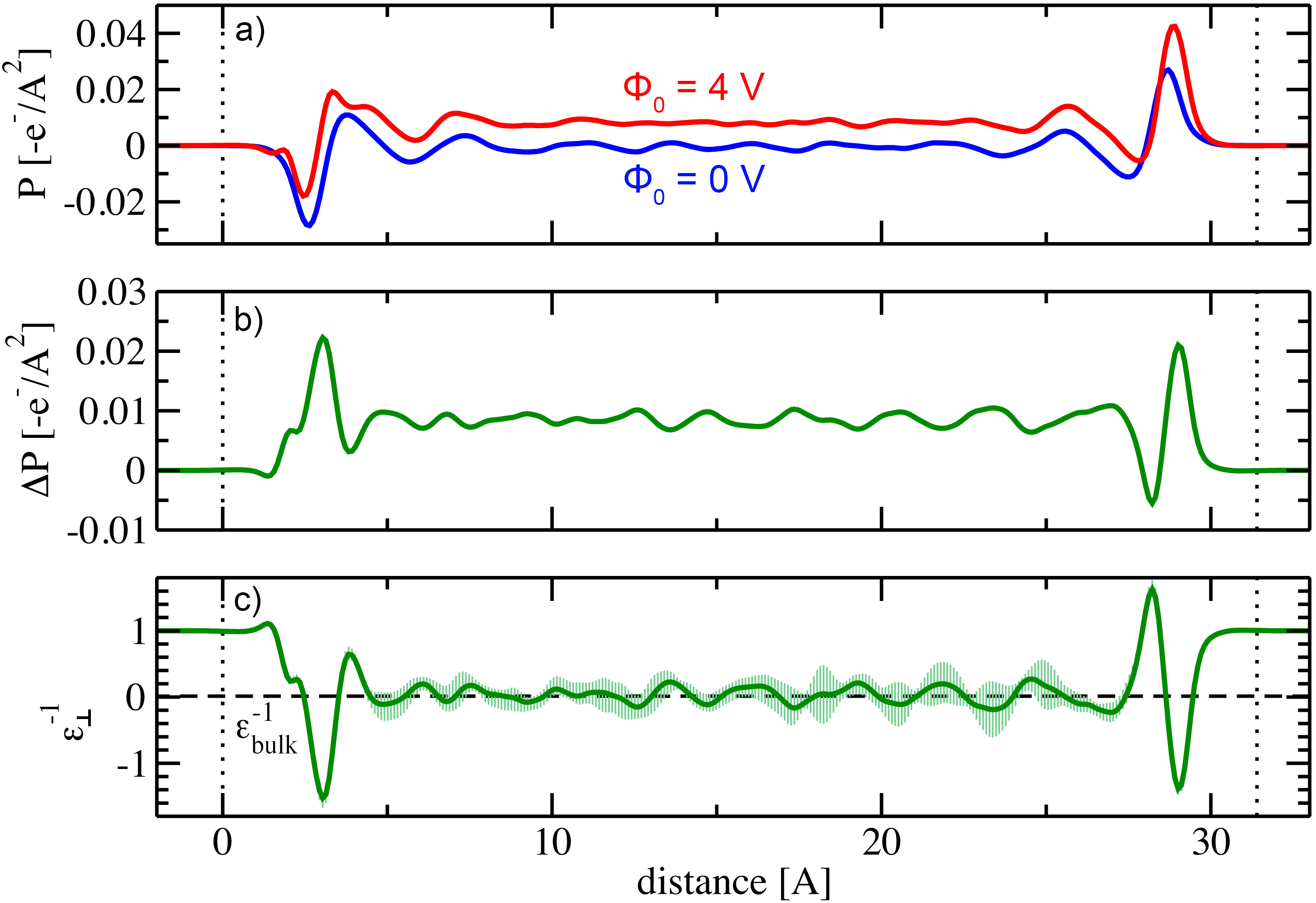}
\caption{\label{inveps}
\textbf{a)} Polarization densities obtained for $\Phi_0 = 0$ V and $\Phi_0 = 4$ V, respectively, and \textbf{b)} polarization density difference. \textbf{c)} Inverse dielectric profile computed from the polarization density difference shown in \textbf{b)} with error bars. The dashed line marks the value of $\epsilon_{bulk}^{-1}$ for liquid water. Dotted lines denote the positions of the electrodes.}
\end{figure}

Such dielectric profiles can be directly computed from the bound charge densities. As a first step, we compute the polarization density $P(z)=\int_0^z \rho_{bound}(z') dz'$, cf. Fig. \ref{inveps}a. For $\Phi = 0\ \mathrm{V}$, the chemical contribution to the bound charge density due to water stratification is reflected in the polarization density close to the interfaces, but $P(z)$ naturally vanishes inside the bulk-like water region. There is no net polarization in this case. The polarization density obtained for an applied voltage of $\Phi = 4\ \mathrm{V}$, in contrast, clearly displays a constant net polarization. We remind the reader that the bound charge densities, and thus by extension the polarization densities, consist of (i) a chemical contribution due to the presence of the interface, superimposed with (ii) an electrostatic contribution that describes the dielectric response. In order to isolate the dielectric response, we compute the polarization density difference $\Delta P$, cf. Fig. \ref{inveps}b, to cancel out any interface related structuring. In linear response theory, this polarization density difference $\Delta P$ is proportional to the inverse local dielectric constant $\epsilon_{\perp}$ with $\Delta E \approx [\epsilon_0 \epsilon_{\perp} ]^{-1} \Delta D $ \cite{loche}. According to the definition of the electric displacement field $\Delta D$, the electric field $\Delta E$ is given by $\Delta E = \epsilon_0^{-1} [ \Delta D - \Delta P ]$. 

Fig. \ref{inveps}c shows the inverse dielectric profile. Compared to dielectric profiles obtained from empirical force-field approaches, which often involve statistical sampling of several nanoseconds, the present simulations were sampled for \mbox{125 ps} and thus exhibit somewhat larger errors bars, indicated by vertical lines in Fig. \ref{inveps}c. The largest error bars by far are encountered inside the bulk-like water region. Close to the interfaces, in contrast, statistical fluctuations are much less pronounced and the interfacial dielectric features shown in Fig. \ref{inveps}c are robust within the numerical accuracy of our simulations.

In comparison to dielectric profiles computed using force-field approaches, cf. e.g. Ref. \cite{deissenbeck}, the first principles dielectric profile presented here features distinct differences between the dielectric responses to the negative and positive charges located on the left and right hand side electrodes, respectively. Moreover,  interfacial water exhibits additional structuring in the dielectric response that is not captured by empirical simulations. Most notably, the inverse dielectric constant assumes values of $\epsilon_{\perp}^{-1} > 1$ directly at the interfaces. The induced polarization of the water therefore locally enhances the applied electric field. This phenomenon is known as antiscreening and was recently proposed to occur within the electrochemical double layer \cite{antiscreening}. Our results demonstrate that antiscreening is already present in neat water.

In conclusion, we extended our thermopotentiostat approach \cite{deissenbeck} towards \emph{ab initio} molecular dynamics (AIMD) simulations and demonstrated its implementation in the context of density-functional theory. We emphasize that the thermopotentiostat can be implemented using any of the currently available finite electric field or charge techniques. In order to highlight the performance of our thermopotentiostat AIMD technique, we computed the dielectric properties of nano-confined water from first principles. These developments allowed us to directly obtain bound charges and polarization densities due to the dielectric response of interfacial water at constant electrode potential. Both the bound charge and the induced polarization density represent important benchmark quantities for future implicit solvent models that are able to accurately describe solvation at the nanoscale. Moreover, our thermopotentiostat AIMD technique is able to describe bond making and breaking, as well as charge transfer processes at electrified solid-liquid interfaces at constant electrode potential. Our developments thus open the door towards simulations of electrochemical and electrocatalytic processes from first principles.

\begin{acknowledgement}
We thank C. Freysoldt, M. Todorova and J. Neugebauer for very helpful discussions. Funding by the German Federal Ministry of Education and Research (BMBF) within the NanoMatFutur programme, Grant No. 13N12972, is gratefully acknowledged.
\end{acknowledgement}

\bibliography{main.bib}

\end{document}





\section{Instantaneously Polarizeable Systems}


Our central working equation (1) in the main paper is equally valid for non-polarizable and instantaneously polarizable systems. As derived in Ref. \cite{deissenbeck}, the fluctuation-dissipation theorem (FDT) takes the following form for the electrode charge:
\begin{eqnarray}
f dt &=& \underbrace{-\frac{1}{\tau_{\Phi}} (\Phi - \Phi_0)\ dt}_{dissipation} + \underbrace{\sqrt{\frac{2}{\tau_{\Phi}} \frac{k_B T}{C_0}}\ dW_t}_{fluctuation}, \label{fdt2}
\end{eqnarray}
where $\tau_{\Phi}$, $\Phi$, $\Phi_0$ and $C_0$ are the relaxation time constant, the instantaneous potential, the target potential and the capacitance of the bare electrodes in absence of a dielectric, respectively. We note that the differential $dW_t$ of a Wiener process plays the same conceptual role as $dt$: it represents an integration over time $t$, albeit with an infinitesimal stochastic time step $dW_t$ using It$\bar{\mathrm{o}}$ integration \cite{gardiner}, cf. Ref. \cite{deissenbeck}.

If the system under investigation is instantaneously polarizable, e.g. in the context of Born-Oppenheimer dynamics, the form of Eq. \ref{fdt2} seemingly suggests that the capacitance $C_0$ would need to be corrected accordingly. Indeed it is possible, in principle, to describe instantaneously polarizable systems by adapting $C_0$ within the fluctuation term. However, this particular choice is inadvisable because it would require advance knowledge about the system's dielectric properties. Instead, the dielectric properties should be an outcome of the simulation, not a parameter entering it. We therefore propose to shift the issue of the unknown dielectric contributions to the capacitance into the time domain and to take into account \emph{any} polarizability, instantaneous or not, \emph{implicitly} within the deterministic term, as described in the following.

For simplicity, we set $\Phi_0 = 0$ without loss of generality and demonstrate the derivation of Eq. \ref{fdt2} explicitly for instantaneously polarizable systems. According to Ohm's Law and Kirchhoff's 2nd Law, the instantaneous current for the setup shown in Fig. 1 in the main manuscript is described by:

\begin{equation}
\frac{dn}{dt} = -\frac{\Phi}{R}, \label{ohmslaw}
\end{equation}

where $n$ and $R$ are the electrode charge and an effective resistance, respectively. We now assume that the capacitance is increased to an unknown value $C = \epsilon_r C_0$, where the factor $\epsilon_r$ describes an instantaneous dielectric response. According to the definition of the capacitance, the instantaneous voltage $\Phi$ is:

\begin{equation}
\Phi = \frac{n}{\epsilon_r C_0}. \label{cap}
\end{equation}

Substituting Eq. \ref{cap} into Eq. \ref{ohmslaw} and adding a corresponding fluctuation term $\tilde{n}\ dW_t$, we obtain:

\begin{equation}
dn = - \frac{1}{R C_0} \frac{n}{\epsilon_r} dt + \tilde{n}\ dW_t. \label{dn}
\end{equation}

In the canonical ensemble at finite temperature $T$, the variance $\sigma_n^2$ of the electrode charge $n$ must satisfy the relation:


\begin{equation}
\sigma_n^2 = k_B T C = k_B T \epsilon_r C_0. \label{ktc}
\end{equation}

Therefore, the fluctuation term $\tilde{n}$ in Eq. \ref{dn} must be constructed accordingly. Here we remind the reader that Eq. \ref{dn} formally represents a stochastic differential equation (SDE) of the so-called Ornstein-Uhlenbeck type:

\begin{equation}
dx = -kxdt + \sqrt D dW_t. \label{OU}
\end{equation}

The variance of Eq. \ref{OU} has been derived analytically using It$\bar{\mathrm{o}}$ calculus \cite{gardiner}:

\begin{equation}
\sigma^2_x = \frac{D}{2k}. \label{OUvar}
\end{equation}

Hence, we directly obtain an expression for $\tilde{n}$, using Eq. \ref{OUvar}:
\begin{eqnarray}
dn &=& \underbrace{- \frac{1}{\tau_{\Phi}} \frac{n}{\epsilon_r}\ dt}_{dissipation} + \underbrace{\sqrt{\frac{2}{\tau_{\Phi}} k_B T C_0}\ dW_t}_{fluctuation}, \label{dn2}
\end{eqnarray}

with $\tau_{\Phi} := R C_0$. Clearly, taking the variance of Eq. \ref{dn2} according to Eq. \ref{OUvar} satisfies Eq. \ref{ktc}. Yet, we note that the fluctuation term is free of $\epsilon_r$. By design, the deterministic dissipation term is the only place where $\epsilon_r$ needs to be introduced.

We remind the reader, that our central working equation (1) in the main manuscript is derived solving the It$\bar{\mathrm{o}}$ integral

\begin{equation}
dn = C_0\ f dt.
\end{equation}

Straightforward multiplication of Eq. \ref{fdt2} with $C_0$ in order to obtain dn and substituting for $\Phi$ using Eq. \ref{cap} yields:

\begin{eqnarray}
dn = C_0\ f dt &=& -\frac{1}{\tau_{\Phi}} C_0 \underbrace{\frac{n}{\epsilon_r C_0}}_{=: \Phi}\ dt + \sqrt{\frac{2}{\tau_{\Phi}} k_B T C_0}\ dW_t, \label{dn3}
\end{eqnarray}

With Eq. \ref{dn3}, it is now clear that Eqs. \ref{fdt2} and \ref{dn2} are formally identical. Our central working equation (1) therefore already takes the polarizability - instantaneous or not - implicitly into account.

\section{Numerical and technical details}

We carried out density functional theory (DFT) calculations within the Perdew-Burke-Ernzerhof generalized gradient approximation \cite{PBE}, using plane wave basis sets and projector augmented wave pseudopotentials \cite{PAW} with an energy cutoff of 400 eV. All calculations were performed using the Vienna Ab Initio Simulation Package (VASP) \cite{vasp1,vasp2}. We used the $\Gamma$ point for $\mathbf{k}$-space integration. To integrate the equations of motion in our AIMD simulations and ensure accurate energy conservation over a time scale of several 100 ps, we converged the electronic total energies to $10^{-8}$ eV at each ionic step, used a discrete time step $\Delta t = 0.5$ fs and the 2nd order leapfrog scheme as implemented in VASP.

The simulation cells contain 2 computational Ne electrodes \cite{surendralal} with a lateral size of \mbox{14.5 $\times$ 14.5 \AA$^2$}, separated by $d = $ 10.7 \AA, 17.4 \AA\ and 31.4 \AA, respectively, and include 32, 64 and 192 H$_2$O molecules between the electrodes, respectively. The values for the electrode separation $d$ were chosen so that the bulk water density of 1 g/cm$^3$ is reached in the central part of the unit cell, after equilibration for 10 ps with the Langevin thermostat and a relaxation time of $\tau = 50$ fs.

From here on, the thermostat was switched off without exception and we sampled the ensembles for an additional 125 ps. In all simulations, we integrated the equations of motion for the spatial degrees of freedom with the leapfrog scheme in the NVE ensemble. In addition to potentiostating the system, the temperature is actively controlled by our thermopotentiostat due to exposing the atoms to the fluctuating electric field, so that the simulation samples the NVT$\Phi$ ensemble.

Both computational Ne electrodes are charged by equal and opposite amounts. The amount of charge transferred between both electrodes is controlled by our thermopotentiostat. To that purpose, we use distinct Ne pseudopotentials for the left-hand side and right-hand side Ne electrodes, respectively. The core charges of the pseudopotentials describing the Ne electrodes are adjusted over the course of the simulation at each individual ionic step, according to our central working equation (1). For the thermopotentiostat relaxation time we use a value of $\tau_{\Phi} = 100$ fs.

\begin{figure}[t]
  \centering
    \includegraphics[width=1\textwidth]{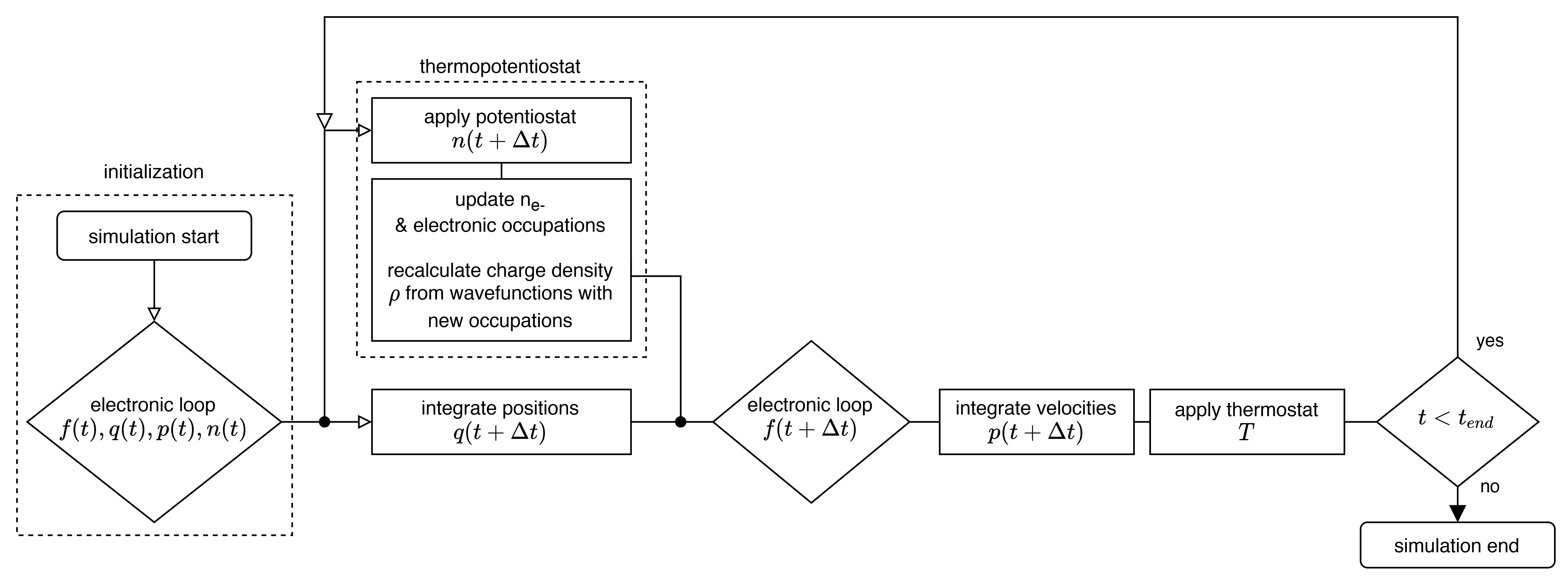}
\caption{\label{integrationscheme} Flowchart to conceptually show the integration of the thermopotentiostat in second order velocity verlet. The potentiostat acts on the charge and positions of the first force calculation and updates the charge according to the thermopotentiostat or any other control logic. The new charge and the new positions are used to calculate the new forces and thereby the new velocities and the next ionic step is performed.}
\end{figure}

\section{Velocity Verlet integration scheme}
In the main text we provided a flowchart for integration via leapfrog. Velocity Verlet is another widely used integration scheme. Here, the thermopotentiostat must be be included in a slightly different way, cf. Fig. \ref{integrationscheme}. After the initialization and a first calculation of the forces the electrode charge is updated together with the positions. Subsequently, a new calculation of the forces is performed which includes the updated charge and updated positions. Next, the velocities are integrated and, if necessary, a thermostat can be applied. The integration loop is then closed by a new integration of the positions and electrode charges.

\section{Convergence of the bound charges and dielectric profiles}
Computing dielectric constants from molecular dynamics simulations is commonly performed using Kirkwood-Fr\"ohlich theory or the theory of polarization fluctuations. Both approaches rely on the variance of the dipole moment fluctuations, typically requiring several nanoseconds of statistical sampling to obtain converged results. Our approach outlined in the main text uses only thermodynamic averages, which converge significantly faster.

\begin{figure}[t]
  \centering
    \includegraphics[width=0.8\textwidth]{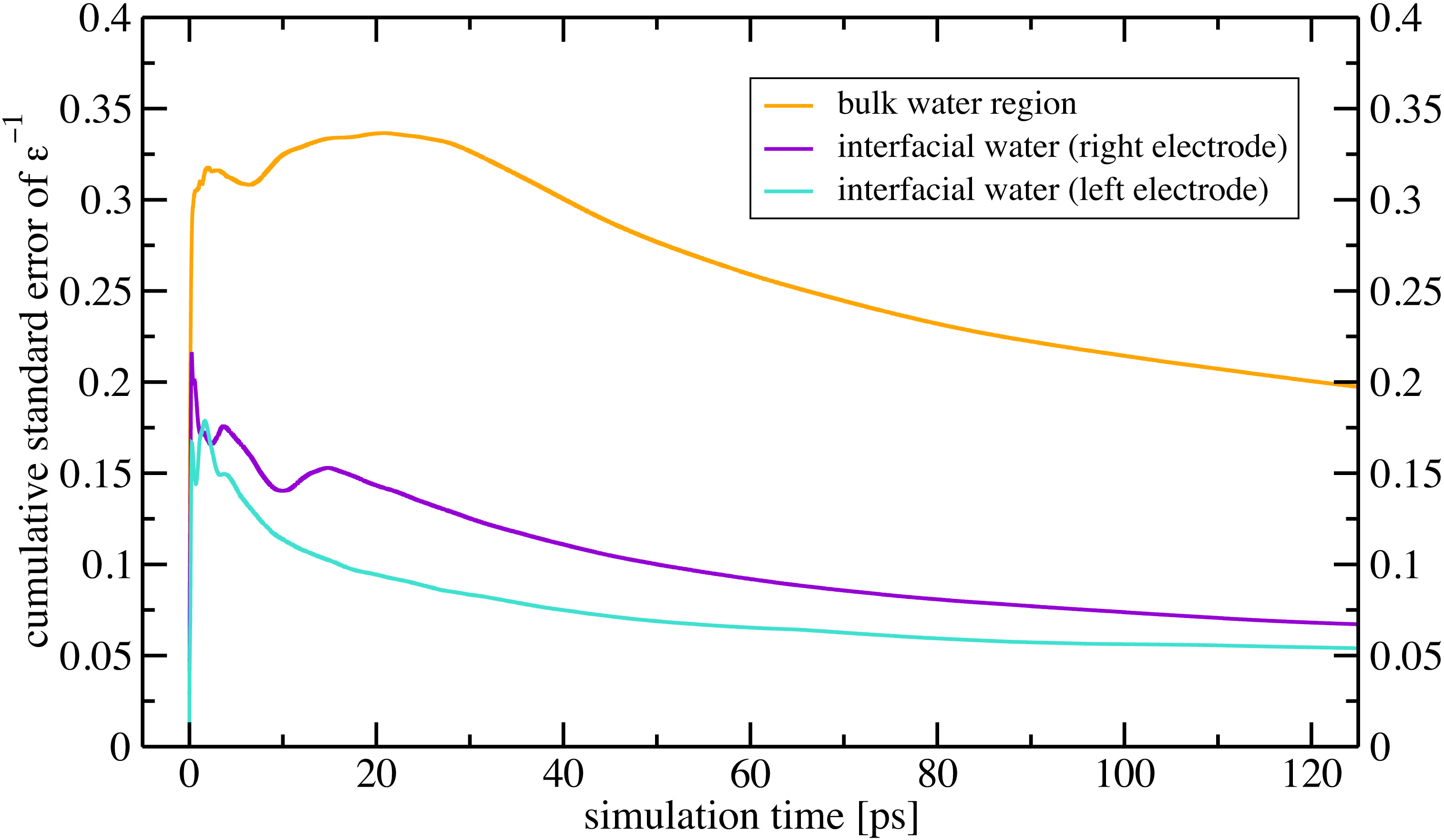}
\caption{\label{variance} Evolution of variances of bulk water dielectric constant with increasing statistics. This plot serves as a guideline to estimate the required number of steps to get a sufficiently accurate dielectric profiles.}
\end{figure}

In order to determine the statistical sampling necessary to converge the dielectric properties, we computed the dielectric constants within the bulk and interfacial water regions as a function of sampling time. Statistical error bars are obtained as running variances:

\begin{equation}
\sigma(z,t) := Var\left( \frac{1}{t} \int_0^t dt' \epsilon_{\perp}(z,t') \right),
\end{equation}
where $\epsilon_{\perp}(z,t')$ denotes the dielectric constant at position $z'$ at timestep $t'$. The position-dependent error bars of $\epsilon(z)$ for the total sampling time are shown in Fig. 6c in the main text. In Fig. \ref{variance} we show the evolution of the error bars as a function of time for the interfacial and bulk water regions. For interfacial water, the standard error of $\epsilon^{-1}$ falls below 0.1 after a sampling time of 50 ps. Consistent with Fig. 6c in the main text, the dielectric properties of interfacial water converge significantly faster compared to those of bulk water, since the water reorientation dynamics is less pronounced close to the interface.

\section{Interfacial water structure}
\begin{figure}[h]
  \centering
    \includegraphics[width=0.75\textwidth]{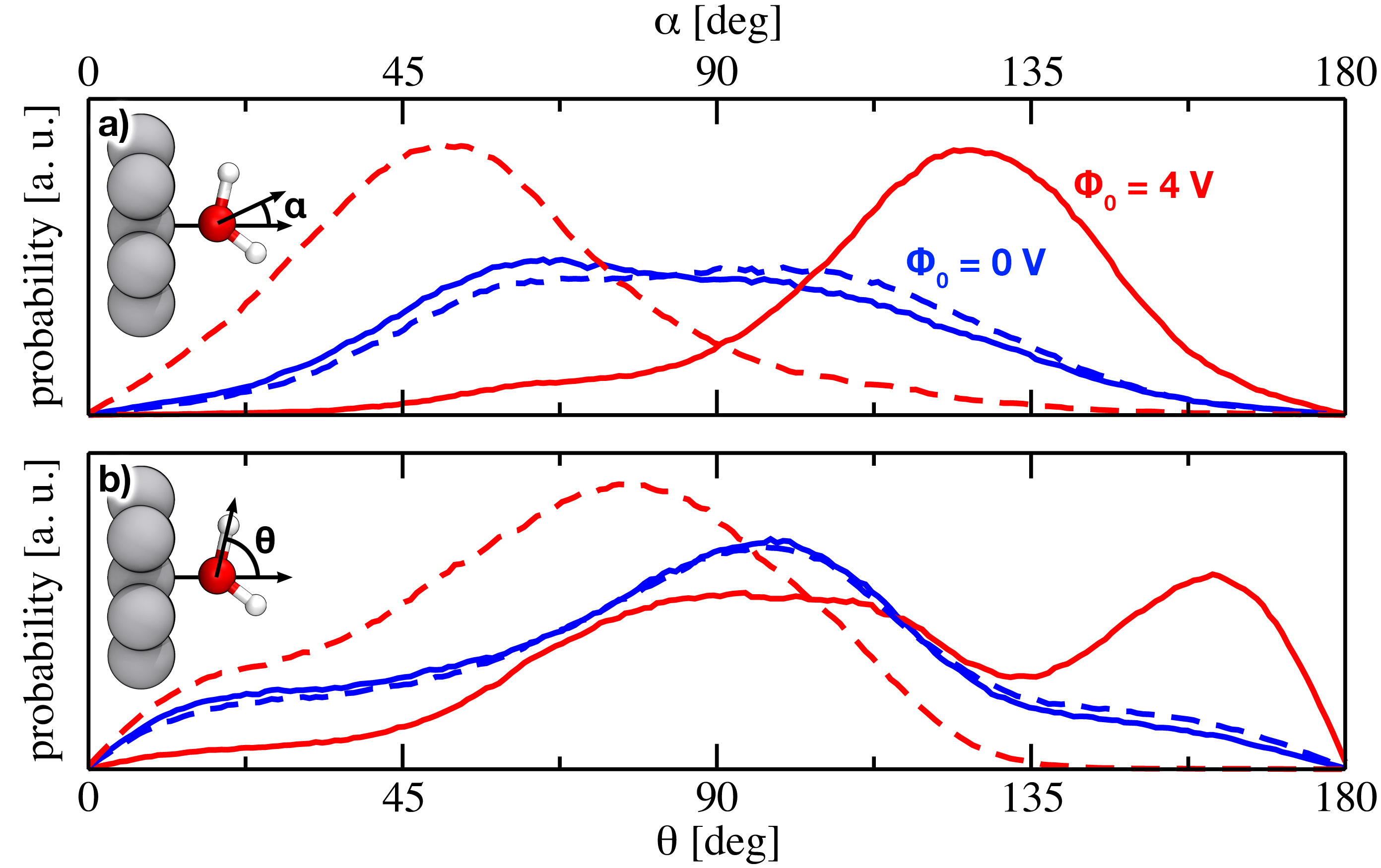}
\caption{\label{orient} Probability distributions of \textbf{a)} the angle $\alpha$ enclosed between the surface normal and the water bisector and \textbf{b)} the angle $\theta$ enclosed between the surface normal and the water OH-bond. Solid and dashed lines indicate distributions computed for negatively and positively charged surfaces, respectively.}
\end{figure}

In order to probe the orientation of interfacial water in response to the applied electric bias, we computed the probability distributions of the angles enclosed between the surface normal and the water bisector ($\alpha$, Fig. \ref{orient}a) or the water OH-bond ($\theta$, Fig. \ref{orient}b). Solid and dashed lines refer to the left hand side (negatively charged) and right hand side (positively charged) electrodes. We consider only the first layer of interfacial water up to a normal distance of 4 \AA\ with respect to the electrode, corresponding to the density minimum between the first and the second stratified water layer, cf. Fig. 5 in the main text.

For $\Phi_0 = 0\ \mathrm{V}$, the angle distributions obtained for the left and right hand side electrodes agree within the numerical accuracy (blue solid and dashed lines, Fig. \ref{orient}), reflecting the symmetry of our computational setup. Both the $\alpha$ and $\theta$ distributions are centered around $90^{\circ}$, indicating that for our hydrophobic electrodes interfacial water adopts largely planar configurations on average, where the molecular planes are parallel to the electrode surfaces.

At an applied voltage of $\Phi_0 = 4\ \mathrm{V}$, we observe a field-induced reorientation of the interfacial water molecules. On the negatively charged left hand electrode (solid red lines, Fig. \ref{orient}) the interfacial water layer features a clear net dipole moment. The maximum of the probability distribution for the angle $\alpha$ between the water bisector and the surface normal is located at $126^{\circ}$. In agreement with recent findings by Li \emph{et al.} \cite{li2019situ} for Au(111) surfaces, the OH-bond angle distribution becomes bimodal: one OH-bond remains in-plane (maximum at $96^{\circ}$), whereas the other OH-bond is now pointing towards the electrode surface (maximum at $162^{\circ}$) in an H-up configuration. On the positively charged right hand electrode (dashed red lines, Fig. \ref{orient}), in contrast, such a bimodal distribution is absent since here the oxygen atoms of the water molecules are oriented towards the electrode surface.

We note that Li \emph{et al.} \cite{li2019situ} used explicit counter ions to induce a surface charge. The interfacial water structure is sampled hence for surface charges that amount to an integer number of electrons and, by extension, for potentials that correspond to those integer charges. The thermopotentiostat approach introduced here, in contrast, allows us to perform simulations under potential control for arbitrary continuous potentials.

\bibliography{main.bib}